\documentclass[a4paper,twoside]{article}

\usepackage{epsfig}
\usepackage{subfigure}
\usepackage{calc}
\usepackage{amssymb}
\usepackage{amstext}
\usepackage{amsmath}
\usepackage{amsthm}
\usepackage{multicol}
\usepackage{pslatex}
\usepackage{apalike}
\usepackage{SCITEPRESS}
\usepackage[small]{caption}

\begin{document}

\title{Training Hybrid Neuro-Fuzzy System  \subtitle{to Infer Permeability in Wells on Maracaibo Lake, Venezuela} }

\author{\authorname{Nuri Hurtado\sup{1}, Raamses D\'iaz\sup{2} and Julio Torres\sup{3}}
\affiliation{\sup{1}Centro de F\'isica Te\'orica y Computacional 
Escuela de F\'isica 
Universidad Central de Venezuela 
Caracas, Venezuela}
\affiliation{\sup{2}S\'ismica BieloVenezolana 
Departamento de Sismolog\'ia 
El Tigre, Venezuela}
\affiliation{\sup{3}Departamento de Ciencias B\'asicas 
Secci\'on de F\'isica 
Universidad Luis Caballero Mej\'ias 
Caracas, Venezuela}
\email{nuri.hurtado@ciens.ucv.ve,diaz\_raamses@hotmail.com,jeta22@gmail.com}
}


\abstract{The high accuracy on inferrring of rockÕs properties, such as permeability ($k$), is a very useful study in the analysis of wells. This has led to development and use of empirical equations like Tixier, Timur, among others. In order to improve the inference of permeability we used a hybrid Neuro-Fuzzy System (NFS). The NFS allowed us to infer permeability of well, from data of porosity ($\phi$) and water saturation ($Sw$). The work was performed with data from wells VCL-1021 ($P21$) and VCL-950 ($P50$), Block III, Maracaibo Lake, Venezuela. 
We trained the NFS with 50\% of data from well $P50$ ($log[k_{P50]}]$, $log[\phi_{P50}]$, $log[Sw_{P50}]$)  to obtain a set of inference equations for permeability 
 ($logk_{P50,i}(log\phi_i,logSw_i)$, $i$ indicates the well to infer). 
These equations were validated with the 50\% of $\phi_{P50}$, $Sw_{P50}$ (the rest of the data not used for training).
 Subsequently, we infer the permeability of the wells $P21$ and $P50$ applying to $k_{P50,i}(\phi_{i},Sw_{i})$ equations the 100\% of $\phi_i$ and $Sw_i$ data of each well ($i=P21\; or\; P50$). 
  We compare the results between empirical Tixier equations  and NFS equations, and we obtain that was possible to improve the permeability inference using NFS's for $P21$ on 21\% and 9\% for $P50$.
We evaluated the NFS equations ($k_{P50,i}(\phi_i,Sw_i)$) with neighboring well data ($P21$), in order to verify the validity of the equations in the area. We have used ANFIS in MatLab.}

\onecolumn \maketitle \normalsize \vfill

\section{\uppercase{Introduction}}
\label{sec:introduction}

\noindent The petrophysical parameters such as capillarity, porosity, permeability, among others, are important  for the charcterization of oil and gas reservoirs.
The permeability is a very complex parameter, its magnitude may change over several orders of magnitude across a reservoir \cite{Finol:2002}. Its estimation from well logs and core analysis is one of the most challenging tasks of a reservoir analyst \cite{Finol:2001}.
Tixier in 1949 \cite{Balan:1995} proposed an empirical equation to calculate the permeability from water saturation ($Sw$) and porosity ($\phi$). 
  Empirical techniques based on well log analysis \cite{Balan:1995,Nelson:1994} and on exponential or power-law techniques that relate permeability with porosity \cite{Shenhav:1971} have also been developed. \\

\noindent Some mathematical approaches apply concepts of either neural networks and/or fuzzy logic to deal with non-linear relationships between two or more variables \cite{Cuddy:2001}. The Neuro Fuzzy System (NFS) method, a hybrid algorithm that combines fuzzy logic with neural networks, has been previously used in the prediction of complex petrophysical parameters \cite{Hurtado:2009,Torres:2007}. In most situations the results obtained have given rise to a set of numerical connections between the different variables involved as well as lithological information about an area of particular interest \cite{Camacho:2013}, paleoclimatic parameters \cite{DaSilva:2010a,DaSilva:2010b}, among others.\\

\noindent In this work, we have compared two techniques for permeability prediction using porosity and water saturation data from two wells at Bolque III, Maracaibo Lake (Venezuela): 
a Neuro-Fuzzy System (NFS) model \cite{Jang:1993,Finol:2001}, and the Tixier relationship \cite{Balan:1995}.

\section{\uppercase{Data and Model}}
\label{model}

\begin{figure}[!t]
\centering
\includegraphics[width=2.5in]{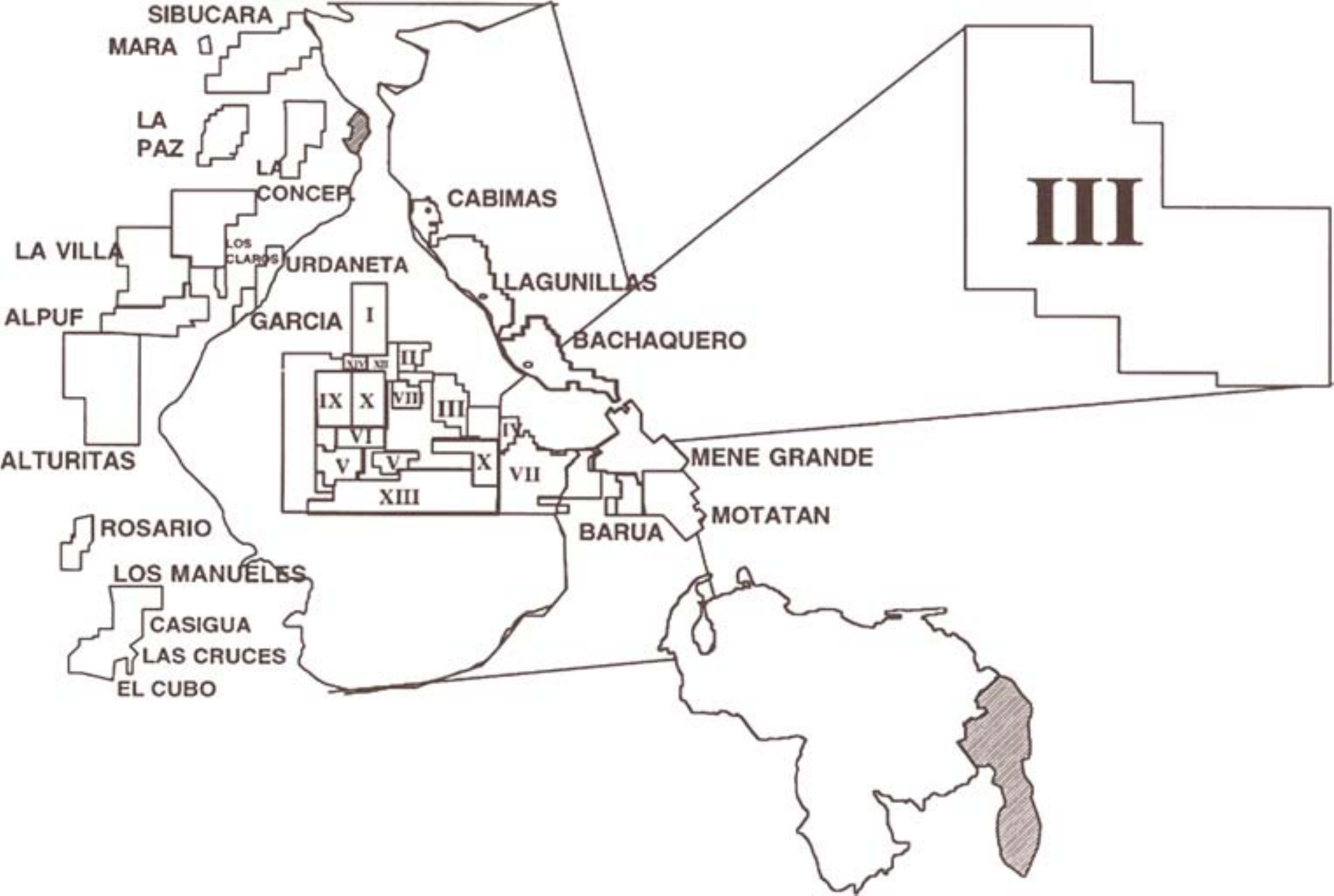}
\caption{Geographical setting of BloqueIII, Maracaibo Lake, Venezuela.}
\label{lake}
\end{figure}

\noindent The data analyzed in this work belong to  Bloque III, Maracaibo Lake, Venezuela (Fig. \ref{lake}). This interval 
comprises units C-455 and C-460 of the Lower Eocene-C. Unit C-455 is a massive sandstone that belongs to the lower 
sequence of interdistributary channels of the area. This unit contains the main accumulation of reserves in the zone. 
Unit C-460 comprises mainly clean thick sands. 
The data of permeability ($k$), porosity ($\phi$) and water saturation ($Sw$) derived of depth interval between 13,200 
and 13,770 ft for the well VLC-950 ($P50$) and between 14,218 and 14,458 ft for the well VLC-1021 ($P21$).\\

\noindent The core porosity values were measured in a porosimeter based on the BoyleÕs-law helium-expansion method. This is a 
standard method for measuring either pore volume or grain volume. It uses BoyleÕs law to determine the pore volume from 
the expansion of a known mass of helium into a calibrated sample holder \cite{Dandekar:2013}. A Gas Permeameter MK-7 
was used to determine the permeability of the samples. This type of permeameter determines the permeability of porous 
solids by forcing a gas, such as air, to flow through the test sample. Measurements of the steady- state flow rate and the 
corresponding pressures provide the necessary data for calculation of the permeability using Darcy law \cite{Dandekar:2013}.
The models we have used in this work to calculate permeability values from porosity core data have different theoretical bases. We are mainly interested in the adaptive neural-fuzzy inference system (ANFIS) \cite{Jang:1993} and empirical  Tixier equation \cite{Balan:1995}.

\subsection{{Neuro-Fuzzy Systems (NFS)}}

\noindent Our NFS is the Adaptative Neuronal Fuzzy Inference System (ANFIS). 
ANFIS is an adaptable hybrid model mixing fuzzy and neural network techniques. 
Also, ANFIS is training from a given input/output data set and adjusts the parameters using a backpropagation algorithm  and one specific membership function.
The training routine for Sugeno-type Fuzzy Inference System consists of a set of fuzzy $if-then$ rules of the form \cite{Jang:1993,Finol:2001}:

\begin{equation}
\label{ifthen}
R_i:\; If\; x_1\; is\; C_{i1}\; and\; x_2\; is\; C_{i2}\; and\; ... 
\; and\; x_n\; is\; C_{in}
\end{equation}
\begin{equation}
Then\; y_i = c_{i1}x_1 + c_{i2}x_2 +...+c_{in}x_n + c_{i0}
\end{equation}

\noindent where the output values $y_i$ are considered as linear or constant functions of the input variables $x_j$. $R_i$ is the ith fuzzy rule; $C_{i1}$,...,
$C_{in}$ are the antecedent linguistic variables and $c_{i1}$, $c_{i2}$, . . . , 
$c_{in}$ the consequent parameters.\\

 The architecture of ANFIS consists of five layers, each layer has a particular objective \cite{Jang:1993}:

\begin{itemize}
\item Layer 1: This layer is composed of n membership functions, each implementing a fuzzy decision rule. Its output is the membership function for which the input variable satisfies the associated $C_{ij}$ term.
\item Layer 2: This layer computes every possible conjunction of the n decision rules.
\item Layer 3: This layer normalizes the conjunctive membership functions in order to perceive the inputs.
\item Layer 4: This layer is a standard perception and associates every membership function with an output (the weights are called consequent parameters).
\item Layer 5: This layer combines all the individual outputs to obtain the total output (sums evidences).
\end{itemize}

\noindent Also, ANFIS supports a TSK system under the following constrains \cite{Jang:1993}:

\begin{itemize}
\item First-order Sugeno-type systems
\item Single output obtained from the weighted average
defuzzification
\item Unity weigh for each rule
\end{itemize}

\subsection{{Empirical Tixier equation}}
\noindent Tixier \cite{Balan:1995}, using empirical relationships between water saturation, resistivity, and capillarity pressure, developed a method to obtain the permeability through the porosity ($\phi$) 
and the irreducible water saturation ($Swi$), using logarithm form:

\begin{equation}
\label{eqn_example}
logk=6log\phi -2logSwi+2log(250)
\end{equation}

\section{\uppercase{Methodology}}

\noindent

\noindent The NFS was implemented using ANFIS of MatLab and its toolboxes.
We trained the NFS with 50\% core data, randomly taken from well $P50$ data ($log[k_{P50]}]$, $log[\phi_{P50}]$, $log[Sw_{P50}]$)  to obtain a set of inference equations for permeability 
 ($logk_{P50,i}(log\phi_i,logSw_i)$, $i$ indicates the well to infer). 
These equations were validated with the 50\% of $\phi_{P50}$, $Sw_{P50}$ (the rest of the data not used for training).\\

\noindent In this work, linear, triangular, bell, pi, and Gaussian membership functions were tested. 
Also, many Fuzzy rules were used: 
$[log\phi_{P50} \; logSw_{P50}] = \{[2\; 1]; [1\; 2];[3\; 1];[1\; 3];[2\; 2]\}$. 
The NFS was trained with $0.1$ of tolerance, and $100$ epochs.
The hybrid optimization method (which is a combination
of least-squares estimation and back propagation) was performed.
 We introduced the data in both, logarithmic and linear forms. 
  For the variables in linear form, eight models were tested:
  
\begin{itemize}
\item Models for $k$
\begin{itemize}
\item A1: $k=a\phi+bSw+c$
\item B1: $k=a\,log\phi+bSw+c$
\item C1: $k=a\phi+b\,logSw+c$
\item D1: $k=a\,log\phi+b\,logSw+c$
\end{itemize}

\item Models for $logk$
\begin{itemize}
\item A2: $k=a\phi+bSw+c$
\item B2:  $logk=a\,log\phi+bSw+c$
\item C2:  $logk=a\phi+b\,logSw+c$
\item D2:  $log k=a\,log\phi+b\,logSw+c$
\end{itemize}
\end{itemize}

\noindent In each case inferred $k$ or $logk$ values were compared with their core data counterparts.
 To quantify the performance of the inference, we estimated the $R^2$ between inferred and core permeability data, and the Root-Mean-Square Error ($RMSE$) values. Finally, the best equations were used to infer permeability for wells $P50$ and $P21$.
 


\section{\uppercase{Results}}

\noindent After adequate number of trials in each case, 
the best inference with the NFS was always accomplished by model D2.
 The table \ref{tabla1}, shows the input/output ranges and best equations  for well $P50$.
  The training was done unsung with two fuzzy rules, R:$[2\;1]$, and gaussian membership function (gaussmf).
  The porosity coefficients in the equations are positive. This behavior corresponds to an increase in the permeability with the porosity that is in agreement with the physical interpretation of the relationship between permeability and porosity given by the KC model \cite{Balan:1995,Finol:2001}.\\

  \begin{table*}
\renewcommand{\arraystretch}{1.3}
\caption{ NFS equations  for well $P50$ and ranges, using  two rules: $[2\;1]$ and gaussian member function (gaussmf).}
\label{tabla1}
\centering
\begin{tabular}{c||c||c}
\hline
\bfseries  Input range & \bfseries Equations  & \bfseries Output\\
$[log\phi]\;\;/\;\;[logSw]$ &   &\bfseries range \\
\hline\hline
 & & \\
 $ [0,51\;1,79]\;\;/\;\;[Inf \;7,75]$ &   $logk_{P50,i}= 0,13log\phi_{i}-0,69logSw_{i}+1,16$    &  \\
                                          &                                                                     & [1,47 5,82] \\
  $[0,20\;0,46]\;\;/\;\;[Inf \;7,75]$ &   $logk_{P50,i}= 0,20log\phi_{i}-33,76logSw_{i}+21,13$ &  \\
& & \\
 \hline
\end{tabular}
\end{table*}

\begin{figure}[!t]
\centering
\includegraphics[width=2.5in]{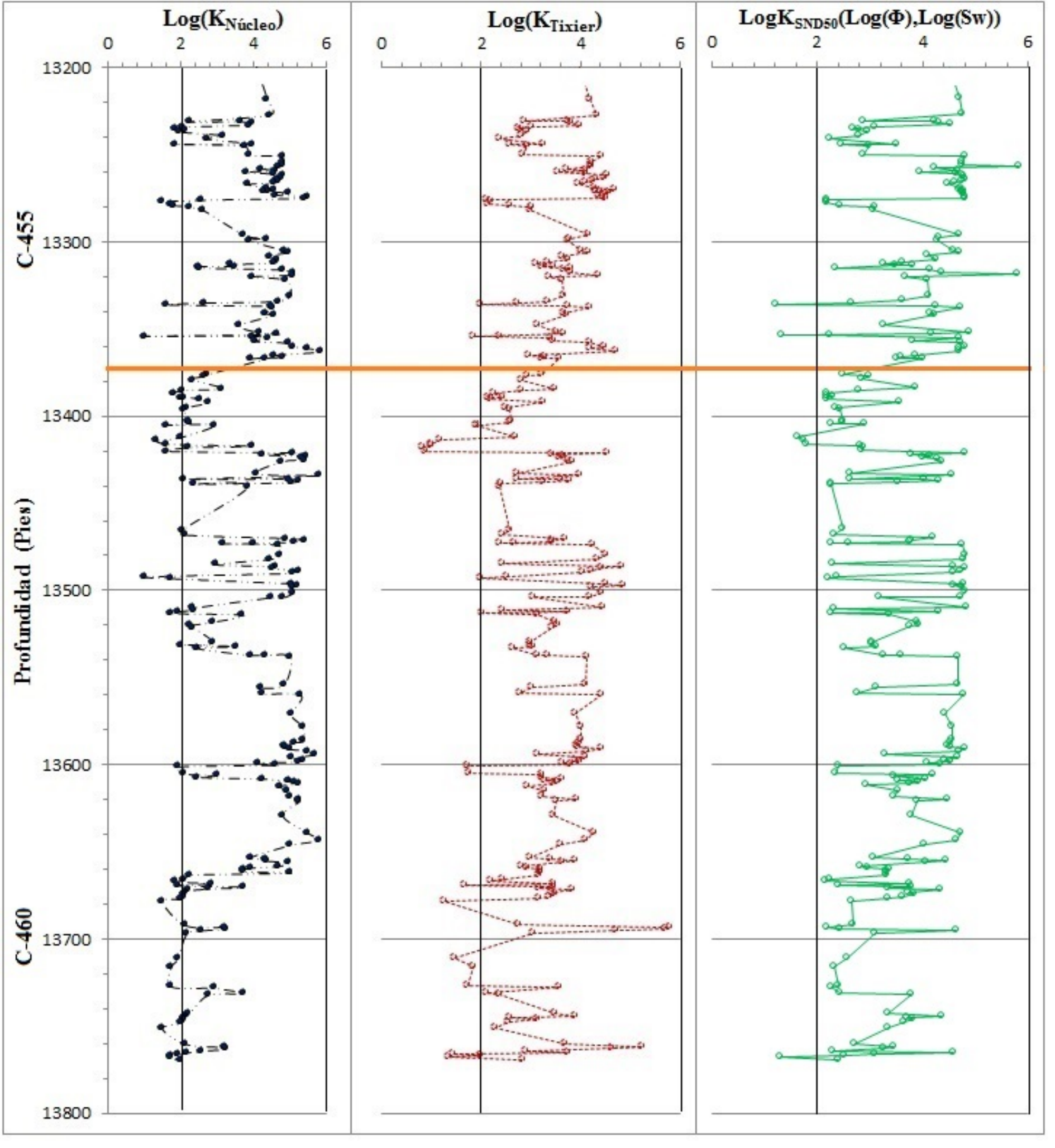}
\caption{Data from well $P50$, logarithm of permeability placed at deep for: (a) permeability core data, $logk_{core}$, (b) inferred empirical Tixier equation, $logk_{Tixier}$, 
and (c) inferred NFS equations, $logk_{P50,P50}$.}
\label{figura2}
\end{figure}

\noindent To quantify the performance of the fitting, we used the $R^2$ correlation between inferred and experimental permeability data, and the Root Mean-Square Error ($RMSE$) values calculated according to:

\begin{equation}
RMSE = \sqrt{\frac{ \sum_{i = 1}^N (k_{inf}-k_{core})^2}{N}}
\end{equation}

\noindent  where $k_{inf}$ and $k_{core}$ are the inferred and core values of permeability respectively, and N is the number of data points. 
In Fig. \ref{figura2} we present the logs of calculated permeability from empirical Tixier 
($Logk_{Tixier}$) and NFS ($Logk_{P50,i}$) models, using 100\% $P50$ data. 
We found that NFS approach has better modeled behavior of the permeability core data qualitatively (Table \ref{tabla2}) and quantitatively (Fig. \ref{figura2}). The modeled was better specially in the shallow zone. This zone is characterized by the presence of cleaner sands, and this could be the explanation for the results obtained.\\

\begin{figure}[!t]
\centering
\includegraphics[width=2.5in]{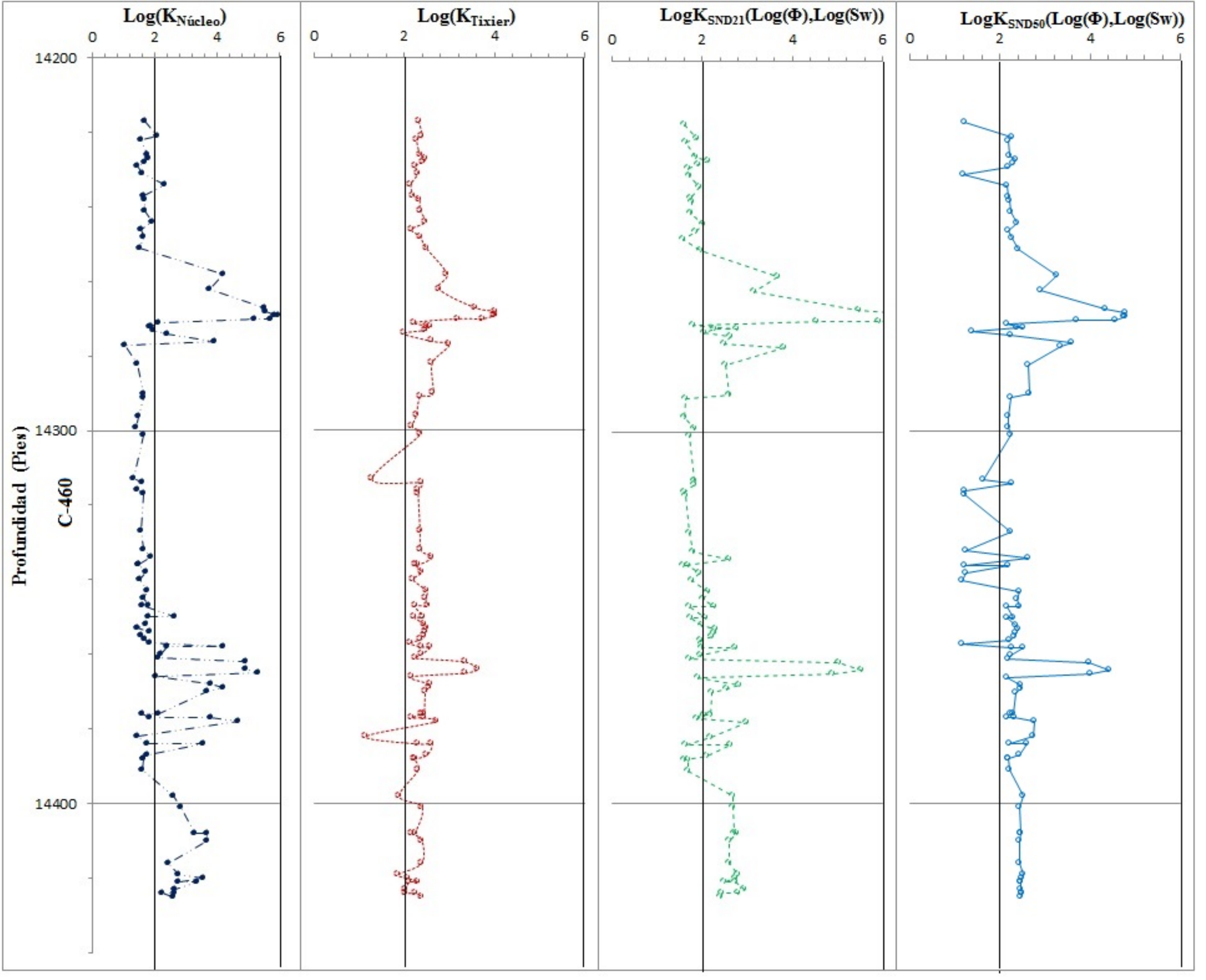}
\caption{Data from well $P21$, logarithm of permeability placed at deep for: (a) permeability core data, $logk_{core}$, (b) inferred empirical Tixier equation, $logk_{Tixier}$, 
and inferred NFS equations obtained training with (c) $P21$ data, $logk_{P21,P21}$, and (d) $P50$ data, $logk_{P50,P21}$.}
\label{figura3}
\end{figure}

\noindent We trained the ANFIS separately with the 50\% of the core  data from wells $P50$ ($logk_{P50,i}$) and $P21$ ($logk_{P21,i}$) and assessed both set of equations with the 100\% of $P21$ core data ($\phi_{P21},Sw_{P21}$). 
The figure \ref{figura3} shows the results of these inferences. 
The qualitative examination of figure \ref{figura3} shows that the results obtained with 
equations: \{$logk_{P50,P21}(\phi_{P21},Sw_{P21})$\}, give a really good inference upon 
permeability core data of $P21$.
We evaluated the NFS equations: \{$logk_{P50,21}(\phi_{21},Sw_{21})$\}, 
with neighboring well data ($P21$), in order to verify the validity of the equations around the area. \\

\noindent The $RMSE$ and $R^2$ values obtained between permeability core data and inference permeability values (Table \ref{tabla2}) confirm the qualitative observation. This table shows the  $RMSE$ and $R^2$ values using:  empirical Tixier equation and NFS equations obtained by training with core data of wells $P50$ an $P21$. All this equations were evaluates with core data of wells $P50$ and $P21$, respectively.

  \begin{table*}[!t]
\renewcommand{\arraystretch}{1.3}
\caption{The $RMSE$ and $R^2$ values obtained between permeability core data and inference permeability values using: empirical Tixier equation and NFS equations obtained by training with core data of wells $P50$ an $P21$.}
\label{tabla2}
\centering
\begin{tabular}{c||c||c||c}
\hline
 \bfseries Wells Data & \bfseries Tixier  & \bfseries $logk_{P50,i}$ & \bfseries $logk_{P21,i}$ \\
\bfseries  to infer (i) & \bfseries  ($RMSE\;/\;R^2$) & \bfseries  ($RMSE\;/\;R^2$) & \bfseries  ($RMSE\;/\;R^2$)\\
\hline\hline
 & & \\
 $i=P50$ & 0,99 \;/\; 0,46 & 0,83 \;/\; 0,55 & $----$\\
 & & \\
  $i=P21$ & 0,86 \;/\; 0,50 & 0,73 \;/\; 0,67 & 0,62 \;/\; 0,71 \\
 & & \\
  \hline
\end{tabular}
\end{table*}

\section{\uppercase{Conclusions}}
\label{sec:conclusion}

\noindent In this work we have used the Neuro Fuzzy System to infer permeability ($k$) 
 from porosity ($\phi$) and water saturation ($Sw$) with data from wells VCL-1021 ($P21$) and VCL-950 ($P50$), Block III, Maracaibo Lake, Venezuela.\\
 
\noindent  The results obtained in this work indicate that, for the studied data, the best approach to permeability from porosity and water saturation is obtained with the statistical approach based on the NFS. Tixier approaches do not improve the results obtained with the NFS.\\
  
\noindent The NFS was training with data from well $P50$, after that, we obtained a set of inference equations: {$logk_{P50,i}$\} (Table \ref{tabla1}). These equations were evaluated with nearby wells data ($i = P 21$). The results indicates that is correct to use NFS-equations to infer permeability around the study area if into we have porosity and water saturation data..

\section*{\uppercase{Acknowledgements}}

\noindent The authors would like to thank to CDHC-UCV for support 
via the research project number PG-03-8269-2011

\bibliographystyle{apalike}
{\small
\bibliography{example}}

\end{document}